\documentclass[fleqn,usenatbib]{mnras}

\usepackage{newtxtext,newtxmath}

\usepackage[T1]{fontenc}
\usepackage{ae,aecompl}

\usepackage{graphicx}  
\usepackage{dcolumn}   
\usepackage{bm}        
\usepackage{comment}
\usepackage{amssymb}   
\usepackage{amsmath}
\usepackage{tabularx}
\usepackage{hyperref}
\usepackage{tabulary}
\usepackage[usenames,dvipsnames]{color}
\usepackage[normalem]{ulem}

\title[Binary mergers with near-zero effective spin]{Precessional dynamics of black hole triples: binary mergers with near-zero effective spin}

\author[Antonini et al.]{
Fabio Antonini,$^{1}$\thanks{f.antonini@surrey.ac.uk}
Carl L.\ Rodriguez,$^{2}$
Cristobal Petrovich,$^{3}$
and
Caitlin L.\ Fischer,$^{2}$
\\
$^{1}$Faculty of Engineering and Physical Sciences, University of Surrey, Guildford, Surrey, GU2 7XH,
United Kingdom \\
$^2$ MIT-Kavli Institute for Astrophysics and Space Research, 77 Massachusetts Avenue, 37-664H, Cambridge, MA 02139, USA\\
$^3$Canadian Institute for Theoretical Astrophysics, University
of Toronto, 60 St George Street, ON M5S 3H8, Canada
}

\date{Accepted XXX. Received YYY; in original form ZZZ}

\pubyear{2015}

\begin{document}
\label{firstpage}
\pagerange{\pageref{firstpage}--\pageref{lastpage}}
\maketitle

\begin{abstract}
The binary black hole mergers detected by Advanced LIGO/Virgo have shown no evidence of large black hole spins.  
However, because LIGO/Virgo best measures the effective combination of the two spins along the orbital angular momentum ($\chi_{\rm eff}$), 
it is difficult to distinguish between binaries with slowly-spinning black holes and binaries with spins lying in the orbital plane.  
Here, we 
study the spin dynamics for binaries with a distant  black hole companion.
For spins initially aligned with the orbital angular momentum of the binary, 
we find that $\chi_{\rm eff}$ ``freezes" near zero as the orbit decays through the emission of gravitational waves.
Through a population study, we show that this  process predominantly leads  to merging black hole binaries 
with near-zero $\chi_{\rm eff}$.
{ We conclude that if the detected black hole binaries were formed in triples, then this
 would explain their low  $\chi_{\rm eff}$ without the need to invoke near-zero 
 spins or initially large spin-orbit angles.}
\end{abstract}




\section{Introduction}
\label{sec:intro}
Since the first detection of merging black hole (BH)
binaries, there has been a proliferation of astrophysical models for producing such systems.  To narrow the range of possibilities, it has been shown that the misalignment between the binary's orbital angular momentum and the BH spins can serve as 
 an important discriminant
between these different formation channels
(e.g., \cite{Rodriguez2016c,2017arXiv170907896F,2017Natur.548..426F}).
Here, we consider binary BH mergers formed through the evolution of stellar triples in the field. In this scenario, the binary is driven to merger
 by the presence of a third BH companion. 
This dynamical configuration can induce high eccentricities in the inner BH binary via the Lidov-Kozai (LK) mechanism \citep{1962AJ.....67..591K,1962P&SS....9..719L}, eliminating the need for the common-envelope phase often invoked in standard binary evolution models 
 \citep{Antonini2014,2017ApJ...836...39S,2017ApJ...841...77A}.
We specifically consider the secular evolution of the 
 effective spin parameter, $\chi_{\rm eff}$  (see Eq.~\ref{chie}),
the combination of BH spins best measured by current 
 gravitational-wave (GW) detectors \citep{2016PhRvL.116x1102A,2016PhRvX...6d1015A,2017PhRvL.119n1101A}.

To date, the BH binaries detected by LIGO/VIRGO have all exhibited small $\chi_{\rm eff}$, with all but one--GW151226--being consistent with $\chi_{\rm{eff}} = 0$ \citep{2016PhRvX...6d1015A,2017PhRvL.118v1101A,2017PhRvL.119n1101A}.
The individual component spins in isolated field binaries are expected to be nearly perpendicular to the binary orbital plane \citep{Kalogera2000}, meaning that current measurements would imply slow rotation of the BHs \citep[if these are formed in the field;][]{2017arXiv170607053B}.
Low spins, however, are in contrast with current observational 
constraints and theoretical expectations, both favouring
rapid BH rotation at formation \citep{2004ApJ...602..312G,2011ApJ...731L...5M}.
In this letter, we show that a more  consistent explanation is possible:
the small values of $\chi_{\rm eff}$ are a consequence of
the spin-orbit tilt produced by the binary's long-term interaction
with a distant companion.

\section{Orbit-averaged Equations}\label{S1}
\label{sec:soss}
We consider a BH binary with total mass $M=m_1+m_2$, 
semi-major axis $a_{\rm }$ and eccentricity $e_{\rm }$, orbited by a tertiary BH with
mass $m_3$ on an outer orbit with semi-major axis $a_{\rm  out}$ and eccentricity $e_{\rm  out}$. We work in terms
of the dimensionless inner-orbit angular-momentum vector, $\boldsymbol{j}={\sqrt{1-e^2}\boldsymbol{\hat{j}}}$, and the eccentricity vector,
${\boldsymbol e}=e\boldsymbol{\hat{e}}$, defined in Jacobi coordinates.
We define the circular angular momenta for the inner and outer orbits as $L_{\rm }=\mu_{\rm }\sqrt{GMa_{\rm }}$ and 
$L_{\rm  out}=\mu_{\rm  out}\sqrt{G(M+m_3)a_{\rm  out}}$ respectively, with 
$\mu_{\rm }=m_1m_2/M$, and $\mu_{\rm  out}=Mm_3/(M+m_3)$.
Finally, the total angular momentum of
the triple system is $\boldsymbol{J}=
L_{\rm }\boldsymbol{j}_{\rm}+
L_{\rm  out}\boldsymbol{j}_{\rm  out}$,
where $\boldsymbol{j}_{\rm  out}$ 
 is the outer-orbit angular-momentum vector. In the absence of dissipation, $\boldsymbol{J}$ is constant.

{ 
A  tertiary  companion highly  inclined with respect to the binary orbit by an angle, denoted by $I$,
 can induce
large amplitude, periodic oscillations in the inner binary eccentricity  \citep[see ][and references therein]{2016ARA&A..54..441N}.
We describe this evolution  using the secular equations at the octupole level of approximation \citep{2015MNRAS.447..747L,2015ApJ...799...27P}.
We also add the 1 and 2.5 post-Newtonian (pN) terms, describing the (Schwarzschild) precession of the argument of 
periapsis 
and  the  orbital decay due to gravitational-wave emission respectively \citep{Peters1964}. The
evolution of the inner binary orbit is determined by the set of equations:
\begin{equation}\label{eqn:1pne1}
\frac{d\boldsymbol{e}}{dt}
= \frac{d\boldsymbol{e}}{dt}\Big|_{\rm LK}+ \frac{3GM}{c^2 a j^3}\nu \boldsymbol{{j}}\times \boldsymbol{e} 
-\frac{304}{15} \frac{G^3\, m_1\, m_2M}{c^5\,a^4
  j^{5}} \left(1+\frac{121}{304}e^2\right) \boldsymbol{{e}} 
\end{equation}
\begin{equation}\label{eqn:1pne1}
\frac{d\boldsymbol{j}}{dt}
= \frac{d\boldsymbol{j}}{dt}\Big|_{\rm LK}+
\frac{304}{15} \frac{G^3\, m_1\, m_2M}{c^5\,a^4
  j^{7}} \left(e^2+\frac{121}{304}e^4\right) \boldsymbol{{j}} 
\end{equation}
  \begin{align}
 \frac{da}{dt} =
 -\frac{64}{5} \frac{G^3\, m_1\, m_2 M}{c^5\, a^3  j^{7}} \left(1+\frac{73}{24}e^{2}+\frac{37}{96}e^{4}\right) \ 
\label{dedtpm}
\end{align}
 where  $\nu=\sqrt{GM/a^3}$ and the LK terms are given explicitly by Eq.~(17)-(20) in \cite{2015MNRAS.447..747L}.
 The associated timescale of the LK oscillations is  \citep[e.g.,][]{2015MNRAS.452.3610A}
\begin{equation}
t_{\rm LK}\approx {1 \over \nu} \left(M\over m_3 \right)\left(a_{\rm  out} j_{\rm  out}\over a\right)^3\ .
\end{equation}
Finally, we add the spin-orbit 
interaction terms \citep[e.g.,][]{Schnittman2004}:
\begin{align}
\frac{d\boldsymbol{S}_{1(2)}}{dt}
&= \frac{2 G \mu}{c^2 aj^{3}} \left(1+\frac{3m_{2(1)}}{4m_{1(2)}}\right)\nu\boldsymbol{j}\times \boldsymbol{S}_{1(2)}   \label{SO3} 
\end{align}
with $\boldsymbol{S}_1$ and $\boldsymbol{S}_2$ the spins of $m_1$ and $m_2$ respectively.
The spin vectors can also be written in terms
of the dimensionless spin parameter $\boldsymbol{\chi}$ as
$\boldsymbol{S}_{1(2)}=\boldsymbol{\chi}_{1(2)} Gm_{1(2)}^2/c$, with $|\boldsymbol{\chi}_{1(2)}|\le 1$.  
 We do not account for the back-reaction torque from $\boldsymbol{S}$ on $\boldsymbol{L}$, as well as the spin-spin precessional terms.
Because these terms depend on the first power of the spin-angular momentum and $jL\gg S$  during the LK oscillations,
they can be safely neglected.}

When describing our results, we will often refer to the evolution of 
the angle between the two spin vectors  and the
spin-orbit misalignment using $\theta_{\rm ss}=\cos^{-1}  \boldsymbol{{\hat S}}_1\cdot \boldsymbol{{\hat S}}_2$ and
 $\theta_{\rm 1(2)}= \cos^{-1}  \boldsymbol{\hat{S}}_{\rm{1(2)}}\cdot\boldsymbol{\hat{j}}$ respectively.  Similarly, the misalignment between the spins and the total angular momentum of the triple is given by $\Theta_{\rm 1(2)}=\cos^{-1}  \boldsymbol{\hat{S}}_{\rm{1(2)}}\cdot \boldsymbol{\hat{J}}$.
Finally, we define the binary effective spin parameter as 
 \begin{equation}\label{chie}
 \chi_{\rm eff}={(m_1 \chi_1 \cos \theta_{\rm 1}+m_2 \chi_2 \cos \theta_{\rm 2})\over M} \ .
 \end{equation}

In the following sections, unless otherwise specified, we work under the  assumption that 
 $\boldsymbol{S}_{1}$, $\boldsymbol{S}_{2}$ and $\boldsymbol{j}$ are initially nearly aligned with each other.
Spin-orbit alignment is generally observed in solitary (solar-type) binaries with relatively 
large separation \cite[$\lesssim 40\rm AU$; ][]{1994AJ....107..306H}
and could either be primordial in origin \citep{2017NatAs...1E..64C} or produced through tidal evolution and mass transfer during stellar evolution prior to BH formation \citep[e.g.,][]{Kalogera2000}.
We note that while a significant spin-orbit
tilt has been measured in a number of cases
\citep{2013A&A...549A..18T,2013ApJ...767...32A,2014ApJ...785...83A}, such misalignment 
is typically attributed to long-term triple evolution  \citep[e.g.,][]{2014ApJ...793..137N}.

\section{Suppression of chaos}\label{precession}
To the lowest pN order, 
each of the two BH spin vectors
precess in response to  torques from the
binary
at the orbit-averaged rate \cite[e.g.,][]{Apostolatos1994}:
\begin{equation}\label{spinp}
\Omega_{\rm S_{1(2)}}= \frac{2 G \mu}{c^2 a j^2} \left(1+\frac{3m_{2(1)}}{4m_{1(2)}}\right) \nu \ .
\end{equation}
\noindent
For fixed $\boldsymbol{j}$, the precession described by Eq.\ (\ref{SO3}) has the
form of uniform precession of the spin vector about the binary angular momentum vector.
\noindent
At the same time, during the LK oscillations, the binary angular momentum vector precesses around the 
total angular momentum of the triple system at a  rate:
$\Omega_{\rm LK}\approx {t_{\rm LK}^{-1}}\ .$

 \begin{figure}
 \includegraphics[width=3.3in,angle=0.]{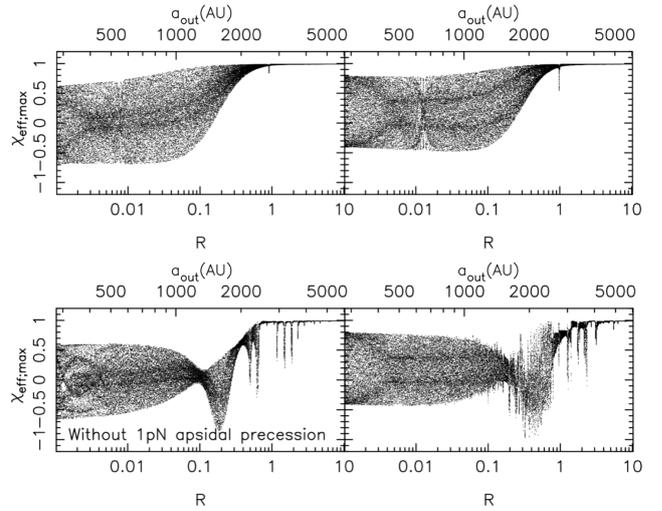}
  \caption{Bifurcation diagrams illustrating the value of $\chi_{\rm eff}$ at the moment the  eccentricity of the inner binary
reaches a maximum during LK oscillations, as a function of the adiabaticity parameters $R$. 
In these integrations we did not include the 2.5pN terms, we set $e=e_{\rm  out}=0.01$ and have assumed
maximal spins. In the left panels we set $m_1=m_3=10 M_\odot$, $m_2=13 M_\odot$, $a=100\rm AU$ and $I=95^{\circ}$;
in the right panels we set $m_1=m_3=10 M_\odot$, $m_2=20 M_\odot$, $a=10\rm AU$ and $I=85^{\circ}$.
The high scatter
observed for some values $R\sim 1$ in the bottom panels, indicates chaotic evolution leading to large spin-orbit angles \citep{2017ApJ...846L..11L}. 
However, when
the 1pN periapsis precession is included in the calculation (top panels), the chaotic behaviour 
is suppressed.}
 \label{fig1s}
 \end{figure}
 
 \begin{figure*}
\centering
\includegraphics[scale=.55]{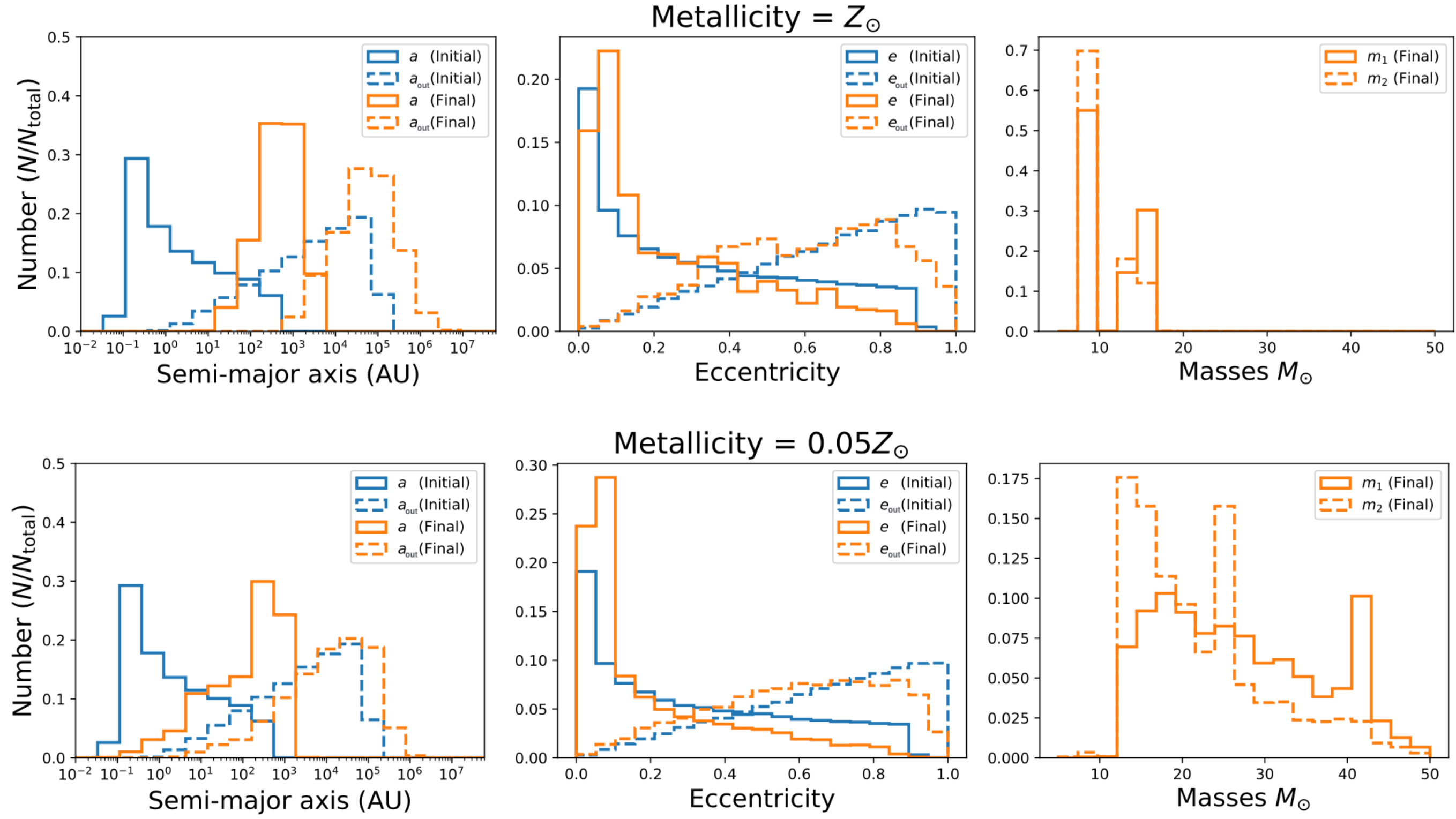}
  \caption{ Examples of the initial conditions from our population synthesis at solar metallicity and for 5\% of solar metallicity. We show the semi-major axes and eccentricities of our initial stellar population in blue and of the evolved BH triples in orange. We only show evolved BH triple systems that remain bound and obey the criteria for evolution stated in the main text (bound, hierarchically stable, and not dominated by Schwarzschild precession).  We also show the mass distribution for the BHs of the inner binaries.  The decreased mass loss at lower metallicities allows for the formation of more massive BHs and lessens the expansion of the inner and outer orbits.
  }
 \label{figIC}
 \end{figure*}
 
We define the adiabaticity parameter: 
\begin{equation}
R_{1(2)}\equiv \left | \frac{\Omega_{\rm S_{1(2)}}}{\Omega_{\rm LK}} \right|_{e=0}=
{2}{G\mu \over c^2a} \left(1+\frac{3m_{2(1)}}{4m_{1(2)}}\right) \nu t_{\rm LK}\ .
\end{equation}
There are three possible regimes for the evolution of each of the two BH spins
 \citep{Storch2014,Storch2015,2017ApJ...846L..11L,2018MNRAS.tmp..135L}:
(i) for $R_{1(2)}\gg 1$ (adiabatic regime),
the spin follows $\boldsymbol{\hat{j}}$ 
 adiabatically, maintaining an approximately constant 
spin-orbit alignment angle $\theta_{\rm 1(2)}$ and consequently a constant
$\chi_{\rm eff}$;
(ii) for $R_{1(2)} \ll 1$ (non-adiabatic regime),  $\boldsymbol{\hat{S}}_{1(2)}$ 
 effectively precesses  about 
 $\boldsymbol{\hat{J}}_{\rm}$,
maintaining an approximately 
constant angle  $\Theta_{\rm1(2)}$; and
(iii) for $R_{1(2)} \approx 1$ (trans-adiabatic regime), 
the spin precession rate
matches  the orbital precession rate and the evolution of 
the spin orbit orientation can  become more complicated.

We might expect that at $R\approx 1$
the spin-orbit angle  will exhibit chaotic evolution due to overlapping resonances,
possibly leading to a wide range of final spin-orbit angles \citep{2017ApJ...846L..11L}. 
As discussed next, we find that
 such chaotic behaviour  is 
 suppressed due to the 1pN precession of the periapsis.

To  lowest order,  the Schwarzschild contribution is
$\omega_{\rm 1pN}=\frac{3GM}{c^2 a j^2} \nu.$
By setting $\omega_{\rm 1pN}/\pi= |dj/dt|/j\approx t_{\rm LK}^{-1}/j$,
we find the critical angular momentum below which LK oscillations are strongly quenched by relativistic precession:
\begin{equation}\label{R1pn}
j_{\rm GR}= {3G\over \pi  c^2}{M^2\over m_3} \left({a_{\rm  out}j_{\rm  out}\over a} \right)^3{1 \over a }\ .
\end{equation}
At $j\leq j_{\rm GR}$, LK oscillations are damped by the in-plane precession
caused by the 1pN terms.
An approximate  criterion for Schwarzschild precession to fully quench the LK oscillations can be obtained by setting $j_{\rm GR}\ge 1$
in the previous equation, which gives:
\begin{equation}\label{bcrit}
{a_{\rm  out}j_{\rm  out}}\geq \left[ \frac{3c^2m_3a }{4GM^2} \right]^{1/3}  a\ .
\end{equation}
See  also \citet{Blaes2002} and \citet{2017ApJ...836...39S} for similar derivations.
We expect that for systems which satisfy this condition the spin-orbit dynamics  leading
 to large misalignment will also be somewhat suppressed.

Fig.\ \ref{fig1s} shows  bifurcation diagrams giving  for each value of $R$ (or $a_{\rm  out}$) the corresponding 
value of $\chi_{\rm eff}$ at every  eccentricity maximum ($\chi_{\rm eff;max}$)
during 100 LK oscillations. 
When $R$ is near unity, the 1pN apsidal precession terms affect the evolution of $\chi_{\rm eff; max}$
in important ways.
If these  terms are not included in our calculations (bottom panels), $\chi_{\rm eff; max}$ 
can attain large values and its evolution is often chaotic
as illustrated by the  high degree of scatter  for a single value of $R$.
Such chaotic behaviour is well known \citep{Storch2014,Storch2015,Anderson2016FormationBinaries}, and 
corresponds to the trans-adiabatic regime discussed above.
However, when all 1pN  terms are  added, $\chi_{\rm eff; max}$  remains close to unity 
at all $R\gtrsim 0.1$.  
We conclude that  the chaotic spin-orbit  dynamics  seen in the bottom panels,
is effectively suppressed  by the  1pN relativistic precession of the inner BH binary orbit.
This is a consequence of the fact that $\Omega_{\rm S}\approx \omega_{\rm 1pN}$
 for any  $a$ and $j$, so that when
 $R\approx 1$  the inequality  Eq.~(\ref{bcrit}) is also satisfied. 
 In this situation, the LK oscillations are quenched
 and only modest misalignment can be induced.

 The results displayed in  Fig.\ \ref{fig1s} indicate (and the numerical simulations below confirm)  that
only when  $R\ll 1$   the spin-orbit orientation does change significantly while
the binary simultaneously experiences extreme eccentricity excitation that can lead to its coalescence through energy loss by GW radiation.

\section{Population synthesis model}\label{pops}
In this section we derive the spin-orbit misalignment of binary BHs driven to a merger
by a distant BH companion using a population synthesis approach.  
We evolved massive stellar triples to BH triples using
a modified form of the Binary Stellar Evolution (BSE)
package \citep{Hurley2002}. 
The tertiary star was  evolved 
simultaneously using the single stellar evolution subset of BSE.

We employed 11 different stellar metallicities, logarithmically-spaced from $1.5 Z_{\odot}$
to $0.01Z_{\odot}$. 
We sampled the primary star mass for the inner binary
from a Kroupa initial mass function \citep{Kroupa2003}
 in the range $22\leq m \leq 150\ M_\odot$. The masses of the secondary 
and tertiary stars were assigned by 
assuming flat mass ratio ($m_2/m_1$ and  
$m_3/(m_1+m_2)$) distributions between $0$ and $1$.
We 
 take $N\propto \log \left(P_1/{\rm days}\right)^{-0.55}$ 
 for the inner orbital period,
 and take the outer orbital periods to be flat in log-space  with $a_{\rm out}\le 10^5\rm AU$.
The eccentricities of the inner binary  was drawn from a $N\propto e^{-0.42}$ distribution with $e$ from 0 to 0.9,
while the eccentricity of the outer orbit was drawn from a thermal distribution, $N \propto 2e$.
Our choice of initial conditions is consistent with observations
of nearby young clusters and associations 
\citep[e.g.,][]{Sana2012}.
All the angles defining the triple (arguments of periapsis, longitudes of the ascending node, and the inclination) were drawn from isotropic distributions.

\begin{figure}
\centering
 \includegraphics[width=2.in,trim={1cm 1cm 0 0}]{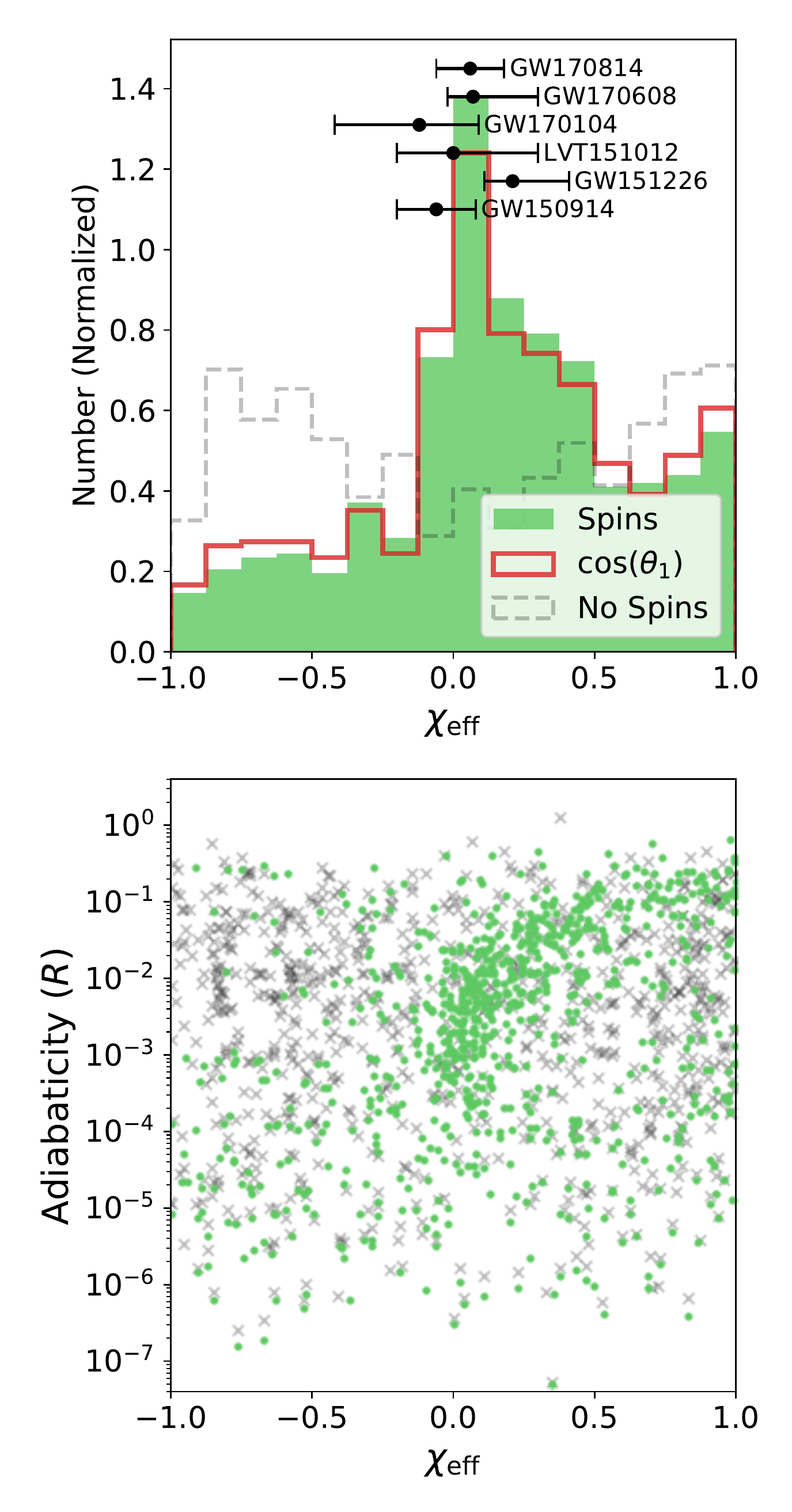}
  \caption{Distribution of final $\chi_{\rm eff}$ for binaries with maximal BH spins
  that merge due to the LK mechanism in our model. 
  The bottom panel shows the value of $\chi_{\rm eff}$ as a function of the 
 adiabaticity parameter described in Section\ \ref{precession}.
 Note the clustering of points near 
$\chi_{\rm eff}\approx 0$ for $R\gtrsim 10^{-3}$ when the spin-orbit terms  (Eq.~\ref{SO3}) are  included
 (green filled circles). 
  The top panel shows the distribution of $\chi_{\rm eff}$ with the inclusion of spin terms (solid green), with no spin terms (dashed grey)
  and the distribution of $\cos \theta_1$ (red).
The  effective spin parameters of the BH binaries detected by Advanced LIGO/Virgo are also displayed.
  }
 \label{fig4}
 \end{figure}

 \begin{figure}
 \includegraphics[width=3.2in,angle=0.]{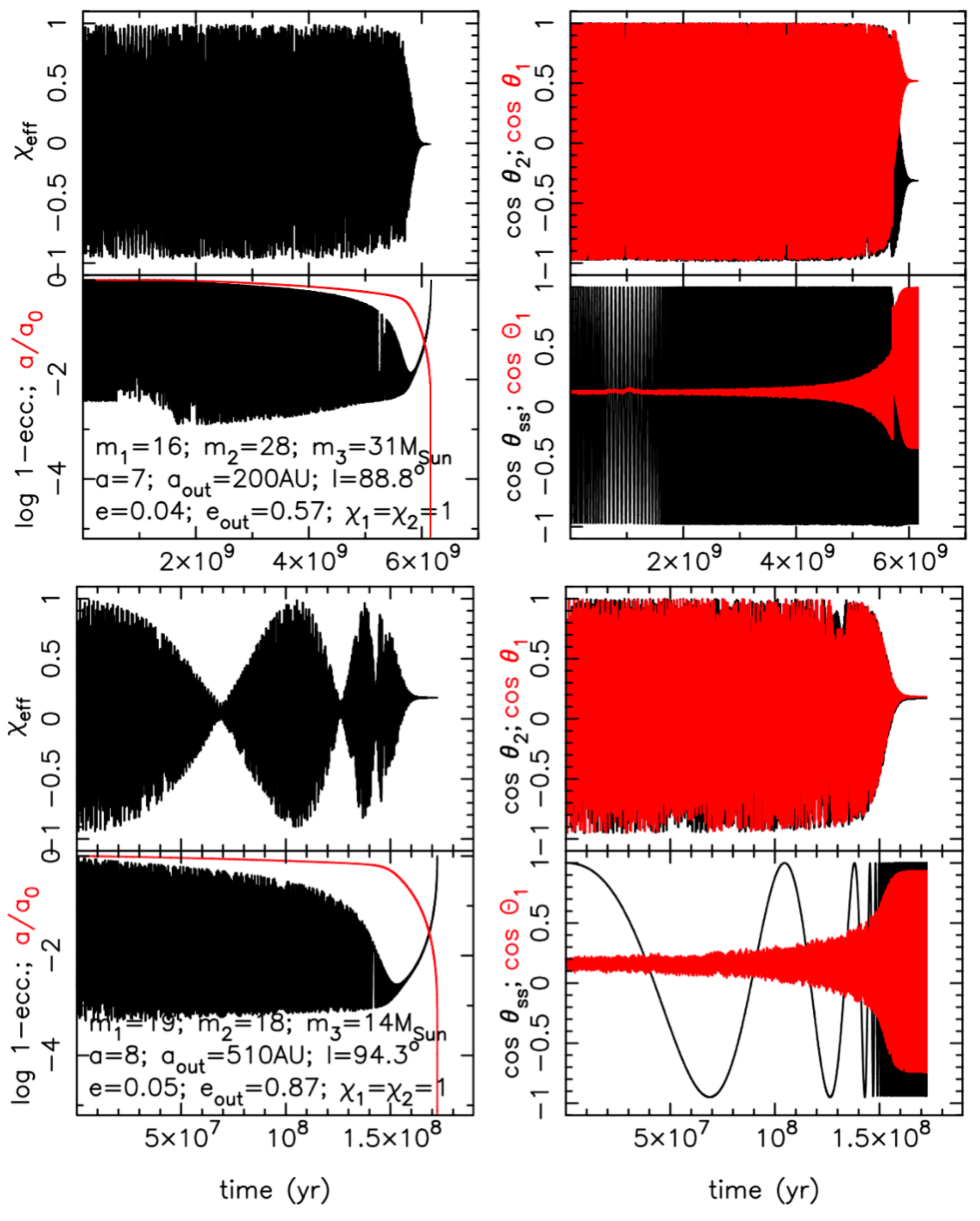}
 \caption{Example cases  of merging systems.
 As the system eccentricity and semi-major axis decrease due to the dissipative 2.5pN terms,
the range of allowed values for $\chi_{\rm eff}$ tends to evolve towards zero.
 Corresponding initial parameters are given in the lower left panels. 
  }
 \label{fig5}
 \end{figure}
 
During the main-sequence  evolution we followed the changes to the initial orbital properties due to the
mass loss and super-nova  kicks  \citep[based on][]{Fryer2012} experienced by each
component of the triple during the formation of a BH \cite[e.g.,][]{2016ComAC...3....6T}.  
Our prescriptions for mass-loss,
stellar winds and natal kicks are identical to
 those in \cite{Rodriguez2016a}, and include the latest prescriptions for the 
 pulsational pair-instability in massive stars \citep{2016AA...594A..97B}.
We reject any systems 
for which either the inner or outer binaries collide or in which the triple becomes secularly unstable at any point \citep{2001MNRAS.321..398M}.

The distributions of initial conditions for the stellar progenitors and for the 
BH triples produced by our models are displayed 
in Fig. \ref{figIC}.
For two metallicities,
we show in these figures the 
distribution of masses, eccentricities and semi-major axes for the bound BH
triple systems that are hierarchically stable, and are not dominated by Schwarzschild precession (i.e., triples with $j<j_{\rm GR}$). 

Our methodology does not consider the possibilities of mass accretion between the inner binary and the tertiary, or any dynamical interaction between the inner and outer binaries. Such physics, while interesting, is significantly 
 beyond the scope of this letter \citep[though  see][ for an analysis of such triples using a self-consistent method]{2017ApJ...841...77A}

Finally, we assume that the  spin vectors of the BHs at the time of their formation are 
aligned with the stellar progenitor spins, and that these progenitor 
spins are initially aligned with $\boldsymbol{{\hat j}}$.  
Although our models include 
the effect of natal kicks on the orientation of the orbits, we find that 97\% (99\%) of our initial population of BH triples have post-kick spin-orbit misalignments less than $0.1^{\circ}$ ($6^{\circ}$).

After generating initial conditions for the BH triples, we integrated 
them forward in time up to a maximum time of $13.8\rm Gyr$, or until the binary peak GW frequency became larger than $10\rm Hz$. 
Fig. \ref{fig4} shows 
the distribution of $\chi_{\rm{eff}}$ for BH binary mergers produced by the LK mechanism in our simulations (which we define
to be those that would not have merged in less than $13.8\rm Gyr$ if evolved as isolated binaries).
In order to obtain the $\chi_{\rm{eff}}$ distribution in Fig. \ref{fig4} we 
 conservatively assumed that both BHs were maximally spinning, but note that
this distribution  can be trivially rescaled to any $ \chi_1=\chi_2$;
 the resulting $\chi_{\rm{eff}}$ distribution is peaked around zero 
 with $\left|\chi_{\rm eff}\right| < \chi_1\times 0.5 (0.25)$ for $\sim 70\%(40\%)$ of the sources. 
The value of $\chi_{\rm eff}$ for the BH binaries detected by Advanced LIGO/VIRGO 
are also displayed in the figure, showing that they are clustered in the range of values
$\left|\chi_{\rm eff}\right| \lesssim 0.5$,  near
the peak of our synthetic distribution. 
The bottom panel of Fig. \ref{fig4} shows that  $\chi_{\rm eff}$ is clustered around
zero 
for initial conditions which are moderately adiabatic, $R\gtrsim 10^{-3}$, demonstrating that the LK process is able to drive 
$\chi_{\rm eff}$ towards zero for a large portion of parameter space for triples. 
{ In our models, most mergers are produced at  metallicities $\lesssim 0.25 Z_{\odot}$. 
In fact, the number of mergers is increased by a factor $\sim 100$ at these lower metallicities primarily due to the reduced
BH natal kicks and stellar mass loss which increase the chance for a triple to remain bound  prior to BH formation.}
{Finally, we find that all our systems become strongly adiabatic (i.e., $R\gg 1$),
after which point  $\chi_{\rm eff}=const.$, before 
the $10$Hz frequency band is reached.
A consequence of this is that all binaries suffer substantial circularisation and 
have $e<0.1$ by the time they enter the LIGO/VIRGO frequency window.}

What is the origin of the near-zero peak of the $\chi_{\rm eff}$ distribution  in Fig. \ref{fig4}?
This peak  could be due to two 
processes: (i) the differential precession of the two spin vectors nearly randomise their relative orientation 
 with respect to each other and with respect to $\boldsymbol{j}$ 
(upper panel in Fig. \ref{fig5});
(ii) both spin-orbit angles evolve individually towards   $\pi/2$ as the orbit slowly decays by 
GW emission (lower panel in Fig. \ref{fig5}).
In the former case, 
$\chi_{\rm eff}$ will peak around zero 
because $\cos \theta_1$ and $\cos \theta_2$ have uniform and independent distributions.
In  case (ii), $\chi_{\rm eff}$ will peak around zero because $\cos \theta_{1}\approx \cos \theta_2\approx 0$. 
Case (ii) turns out to be more important. 
This is shown in Fig. \ref{fig4}, where we see that the final distributions of $\cos\theta_1$  and $\cos\theta_2$ (not shown) follow
closely the distribution of
 $\chi_{\rm eff}$ and they are also peaked around $0$\footnote{The reason
 for this ``attractor'' towards $\pi/2$ has been recently identified in \citet{2018arXiv180503202L}
 while this letter was under review.}.  

\section{Conclusions}
For BH binary mergers produced from isolated field binaries, it is expected
that the individual BH spins should be nearly aligned
with the binary angular momentum \citep{Kalogera2000}.
In these standard population synthesis models,
the relatively small $\chi_{\rm eff}$
of the BH binaries detected so far by LIGO/VIRGO 
\citep{2016PhRvL.116x1102A,2016PhRvX...6d1015A,2017PhRvL.119n1101A}
is more easily explained
if the spin magnitudes were nearly zero.
{  This, however, appears to be disfavored by current
observational constraints and theoretical models which 
suggest finite spins for BHs at birth \citep{2004ApJ...602..312G,2011ApJ...731L...5M}.
A triple origin instead,
could provide a more  consistent explanation for
the low $\chi_{\rm eff}$ values (as shown here), as well as for their merger rates \citep{2018arXiv180508212R}.
We note, however, that to be a viable explanation for {\it all} the LIGO/VIRGO detections, our mechanism 
 would require strong suppression of  other channels.}
\\

\noindent
{\it Software:} the secular code used in this paper 
is available at \href{https://github.com/carlrodriguez/kozai}{https://github.com/carlrodriguez/kozai}.

\noindent
FA acknowledges support from  an STFC E. Rutherford
fellowship (ST/P00492X/1),
CR from a Pappalardo fellowship at MIT,
CP  from the J.~L. Bishop Fellowship and from the 
Gruber Foundation Fellowship, 
CF from a Sir Edward Youde Memorial Fund scholarship and the Undergraduate Research Opportunities Program at MIT.
FA and CR 
 acknowledge the hospitality and support
of the Aspen Center for Physics (NSF Grant PHY-1607611).
\bigskip

\bibliographystyle{mnras}

\bsp	
\label{lastpage}
\end{document}